# A Note on Aryabhata's Principle of Relativity

Abhishek Parakh

**Abstract:** This paper presents the principle of relativity of motion as described in Aryabhata's text *Aryabhatiya*. This principle is likely to have been instrumental in the framing of Aryabhata's theory that the earth rotated on its axis and, therefore, it has played a very important role in the history of astronomy.

## 1 Introduction

It is not generally known that the first clear description of relativity was provided by Aryabhata [1], the great Indian astronomer, who was born in 476 CE. In this note, we present the relevant section from *Gola-pada* (Astronomical Section) in his book *Aryabhatiya*, which proves that Aryabhata was fully aware of the relativity of motion.

Aryabhata, occupies an important place in the history of mathematics and astronomy. He took the earth to rotate on its axis and he gave planet periods with respect to the sun. For some recent papers on Aryabhata's work, see [2-7], and for books, see [8-10]. In [3], Aryabhata's ideas related to the size of the planetary system are discussed, and in [4], the origin of his planetary parameters is discussed. Thurston, in his essay [9], summarizes: "Not only did Aryabhata believe that the earth rotates, but there are glimmerings in his system (and other similar Indian systems) of a possible underlying theory in which the earth (and the planets) orbits the sun, rather than the sun orbiting the earth."

In *Gola-pada,* Aryabhata presents 50 *shlokas* (stanzas) that deal with the motion of the sun, moon and the planets. His text also describes the motion of the celestial sphere as seen by those on the equator and by those on the north and south poles; and gives rules relating to the various problems of spherical astronomy. It also deals with the calculation and graphical representation of the eclipses and the visibility of the planets.

This article discusses his principle of relativity, and we argue that it must have helped him to reach the conclusion that the earth rotated on its axis.

## 2 Relativity in Aryabhatiya

The following *shloka* 9 (stanza 9), from *Gola-pada* (Astronomical Section) of *Aryabhatiya,* presents his principle of relativity in a cryptic form:

अनुलोमगतिनौंस्थ: पश्यत्यचलं विलोमगं यद्वत् ।

अचलानि भानि तद्वत् समपश्चिमगानि लङ्कायाम्   ॥ ९ ॥

**Translation** [1]: *Just as a man in a boat moving sees the stationary objects (on either side of the river) as moving backward, just so are the stationary starts seen by people at Lanka (reference co-ordinate on the equator), as moving exactly towards the west.*

The above statement shows that Aryabhata was aware of the problem associated with absolute velocity. He was aware that motion of an object is only with respect to a frame of reference. Here, Aryabhata has used the boat as one frame of reference and objects on bank of the river as other frame of reference; these are the uniformly moving and stationery frames. Having illustrated his principle using a terrestrial example, he applies it freely to celestial objects, stating that stars are stationary and we are moving. This application of a terrestrial idea to the stars is very significant conceptually, since it means that the same laws should apply everywhere.

*Shloka* 10 (stanza 10):

उदयास्तमयनिमित्तं नित्यं प्रवहेर्ग् वायुना क्षिप्त: ।
लङ्कासमपश्चिमगो भयञ्जर: सग्रहो भ्रमति ॥ १० ॥

**Translation** [1]: *(It so appears as if) the entire structure of the asterisms together with the planets were moving exactly towards the west of Lanka, being constantly driven by the provector wind, to cause their rising and setting.*

The above stanza implies that Aryabhata was perfectly aware that experiments dealing with observations in a different frame of reference with respect to one on which the observer is situated will provide exactly similar results. This conclusion is drawn because as seen in the previous stanza he had described stars as stationary objects while in the present one he says stars appear to move due to the "provector wind." This also means that he saw mechanism in terms of a provector wind as a kind of an artifact.

## 3   Antecedents

For the antecedents to Aryabhata's principle of relativity one must go to the earlier sources on Vedic astronomy [11-15]. That his work was grounded in measurements is established by Billard [16], who shows that his parameters are true for Pataliputra (Patna) of about 510 C.E. when the book appears to have been written.

The Aitareya Brahmana (3.44) says: "The Sun never really sets or rises." Another very early text, the Shatapatha Brahmana (8.7.3.10) states**:** "The sun strings these worlds - the earth, the planets, the atmosphere - to itself on a thread." Both these imply that the matter of representation of motion with respect to earth as well as sun was part of the astronomical tradition.

We conclude that there was a very ancient tradition related to considering the sun as the centre of the solar system in India. Although some naïve ideas of relativity must have been at the basis of this tradition, a clear expression of this thinking is to be found for the first time in India in the work of Aryabhata.

It is possible that this relativity principle led the authors of the medieval Vishnu Purana (2.8) to assert: "The sun is stationed for all time, in the middle of the day. [...] Of the sun, which is always in one and the same place, there is neither setting nor rising." [14,17].

## 4 Galileo's Relativity

It is worthwhile to compare Aryabhata's relativity principle to that of Galileo (1564-1642), who explained his principle of relativity by stating that one cannot use any mechanical experiment to determine absolute constant uniform velocity.

Galileo presented the main idea behind his principle in the *Dialogue Concerning the Two Chief World Systems* (1632). In this work, Salvatius describes two scenarios concerning ship's cabin. In both, two friends are in the cabin, along with butterflies and other small flying animals, fish swimming in a bowl, a bottle from which drops of water fall into another container, and a ball. The cabin is below deck, so neither person can see outside.

In the first scenario, the ship is at rest; in the second, the ship is traveling at a constant velocity. In both, the animals and the fish move about, and the two friends go about various activities, and there is no way they can tell a difference. The motion of the ship has no effect on the activities in the cabin. This is the central idea behind Galileo's principle of relativity.

It is striking that the examples put forward by Aryabhata and Galileo are fundamentally identical. In each case there are two scenarios: one with a frame in motion and the other with the frame at rest. In each case, effectively, the local observations remain unaffected by the motion of the frame.

## 5 Conclusions

Aryabhata's principle of relativity appears to have played a central role in the development of astronomical models in India and, therefore, it is of great interest to the historian of astronomy.

## References

1. K.S. Shukla and K.V. Sarma, Aryabhatiya of Aryabhata. Indian National Science Academy, 1976.